\documentstyle[prl,aps,multicol,psfig,epsfig]{revtex}

\begin{document}
\draft
\title{Charge Transport in a Spin-Polarized 2D Electron System
in Silicon}
\author{D.~A.~Knyazev, O.~E.~Omel'yanovskii,
A.~S.~Dormidontov, and V.~M.~Pudalov}
\address{P.~N.~Lebedev Physical Institute, Leninski
prospekt 53, Moscow, 119991 Russia}
\maketitle
\begin{abstract}
The temperature dependences of the conductivity $\sigma(T)$ for
strongly interacting 2D electron system in silicon have been
analyzed both in zero magnetic field and in spin-polarizing
magnetic field of 14.2\,T, parallel to the sample plane.
Measurements were carried out in a wide temperature range
$(1.4-9)$\,K, in the ballistic regime of electron–-electron
interaction, i.e., for $T\tau > 1$. In zero magnetic field, the
data obtained for $\sigma(T)$
 are quantitatively
described by the theory of interaction corrections. In the fully
spin-polarized state, the measured $\sigma(T)$ dependences are
nonlinear and even nonmonotonic for the same temperature range,
where the $\sigma(T)$ dependences are monotonic in the absence of
the field. Nevertheless, the low-temperature parts of the
experimental $\sigma(T)$ dependences are linear and are
qualitatively consistent with the calculated interaction
corrections.
\end{abstract}
\pacs{PACS: 71.10.Ay, 75.47, 71.30.+h}

\begin{multicols}{2}

Electron-electron interaction causes a number of significant
effects in the conductivity and magnetoconductivity of 2D systems
of charge carriers \cite{review1}: in particular (i) a strong
metallic-like temperature dependence of conductivity
($d\sigma(T)/dT> 0$) in zero magnetic field
\cite{review1,review2,PRB94,PRB95}, and (ii) strong
magnetoresistance in the in-plain magnetic field
 \cite{prl97,jetpl97,physicaB98,okamoto99}.

At present, several models claim to explain these and other
effects of the renormalization of the quasiparticle parameters by
strong electron-electron interaction \cite{review2} both within
the framework of the Fermi-liquid theory and beyond this
framework. In this work, we compare with experimental data
predictions (a) of  the theory of interaction corrections
\cite{ZNA1} and, (b) of the two-phase model \cite{spivak}, where
the 2D electron system is treated as consisting of a electron
liquid with Wigner crystal inclusions. The former model
qualitatively, and in some cases, even quantitatively reproduces
the temperature dependence of the conductivity and
magnetoconductivity of 2D systems
\cite{aleiner,savchenko,shashkin,vitkalov,kvon} as a consequence
of Fermi-liquid effects. The latter model qualitatively explains
the above effects in conductivity from the essentially non
Fermi-liquid viewpoint.

Investigation of the system in a magnetic field
higher than the field of full spin polarization
$B_{pol}$, applied
parallel to the 2D plane, can be a sensitive test for the
above theories. This case is rather simple from the theoretical
point of view (because parallel magnetic
field affects only spins of electrons and does not
affect their orbital motion), whereas the predictions of different
theories for the spin-polarized state significantly differ
from each other. For this reason, studies of electron transport
in the completely spin polarized state
can provide an additional key for understanding
the properties of the strongly interacting 2D electron
system.

Measurements of $\sigma(T)$ in the fully spin-polarized state have
already been performed in Refs.~\cite{sp_vitkalov,sp_shashkin}. In
particular, it was reported in Ref. \cite{sp_vitkalov} that the
conductivity in a magnetic field of 10~T is almost temperature
independent in the temperature range $T= (0.3 - 3)$\,K. This result
seems to be in accord with the predictions of the two-phase model
\cite{spivak}. However, a field of 10\,T was likely to be
insufficiently strong for the complete spin polarization of the
electron system in the studied density range
$(1-2.5)\times10^{11}$\,cm$^{–2}$. The ratio
$\sigma_D^{-1}d\sigma/dT$ of the derivatives measured in zero
magnetic field and in the full spin polarization field was
measured in Ref.~\cite{sp_shashkin} in a temperature range $(0.1-
1.2)$\,K. It was found that this ratio is consistent with the
theory of temperature-dependent screening
\cite{stern,GD86,DG00,DSH_SP}
 and is inconsistent with calculated interaction corrections
\cite{ZNA1} for the ballistic regime. However, it is
worth noting that the applied strong parallel field
increases the effective disorder in the 2D system
(reduces $\tau$). For this reason, although the temperature
range of measurements in zero magnetic field corresponded
to the ballistic interaction regime $T\tau > 1$
in both works \cite{sp_vitkalov,sp_shashkin},
the same temperature range in the
strong magnetic field corresponded to a transient
region $T\tau \sim 1$ between the diffusive and ballistic interaction
regimes. Thus, the comparison of the $\sigma(T)$ slopes  in zero field and in strong field
was performed for two different
regimes of electron–-electron interaction.

In this paper, we report measurements of the temperature
dependence of conductivity performed for higher temperatures,
$(1.4-9)$\,K, and in a stronger magnetic field,  14.2\,T. Under
such conditions, the measurements certainly corresponded to the
ballistic interaction regime $T\tau > 1$ both in zero magnetic
field and in the field of the full spin polarization. The results
of our $\sigma(T)$ measurements in strong magnetic field disagree
with the non Fermi-liquid model \cite{spivak} that predicts
vanishing temperature dependence of conductivity in the
spin-polarized state. Our data for zero magnetic field, as well as
the data in Ref.~\cite{aleiner}, are quantitatively consistent
with the theory of interaction corrections \cite{ZNA1} with no
free parameters. In the spin-polarized state, the temperature
dependence of conductivity is only qualitatively consistent with
calculated interaction corrections, whereas the slopes of the
experimental $\sigma(T)$ curves differ from the theoretically
predicted values by a factor of $1.2-2.8$.

Measurements were performed on a (100)-Si MOSFET structure with
the inverse electron layer, $5 \times 0.8$\,mm$^2$ rectangular
channel, and maximum carrier mobility $\mu \approx 2$\,m$^2$/Vs at
$T = 1.4$\,K. The conductivity was measured by  the standard
four-terminal ac technique  with a frequency $\approx 5$\,Hz. The
current passing through the sample was small enough ($I \approx
10$\,nA) to prevent the overheating of electrons. The sample plane
was parallel to the magnetic field with an accuracy $\approx 1'$.
The orientation was controlled according to vanishing of the weak
localization peak in $\rho_{xx}(B_\perp)$ when the sample plane
rotated with respect to the magnetic field direction.

\subsection{Measurements of $\sigma(T)$ at zero magnetic field}
\begin{figure}[h]
\centerline{\includegraphics[scale=0.9]{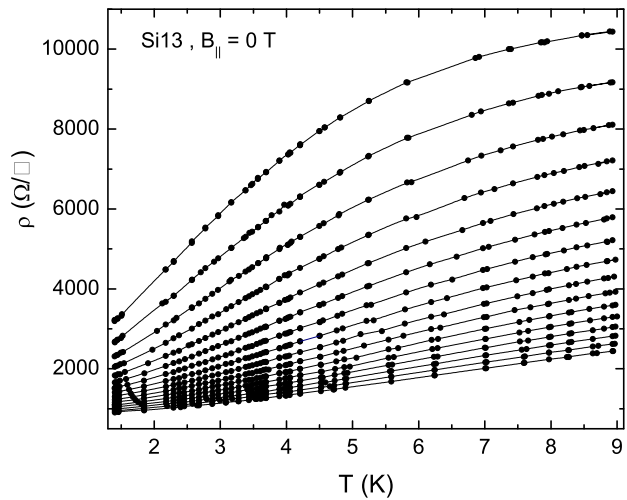}}
\caption{Temperature dependence of $\rho(T)$ for
densities 1.92, 2.04, 2.16, 2.28, 2.39, 2.51, 2.63, 2.75,
2.86, 2.98, 3.10, 3.22, 3.34, 3.45, 3.57, 3.69 (from top to bottom, in  $10^{11}$\,cm$^{-2}$).
Magnetic field $B_\parallel = 0$\,T.}
\label{Fig.1}
\end{figure}

Prior to studying conductivity
 in strong magnetic fields, we have done $\sigma(T,B)$ measurements in zero magnetic
field and compared the results with theory. Figure~\ref{Fig.1}
shows the $\rho(T)$ dependences
 in zero magnetic field for
various electron densities. In the absence of the field,
the  sample resistivity strongly depends  on
temperature (see Fig.~\ref{Fig.1}), exhibiting metallic conductivity
typical for high mobility samples for low carrier
densities (i.e., for strong electron-electron interaction).

\begin{figure}[h]
\centerline{\includegraphics[scale=0.95]{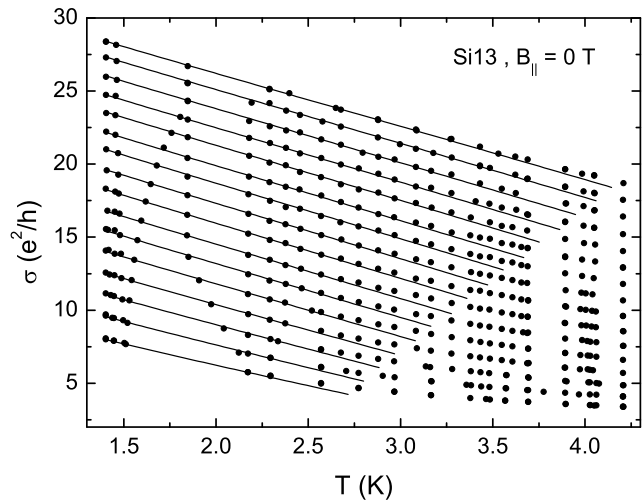}}
\caption{Comparison of the experimental $\sigma(T)$
dependences (dots) and  calculated interaction corrections Eq.~(\protect\ref{sigTB0})
(solid lines) for densities
1.92, 2.04, 2.16, 2.28, 2.39, 2.51, 2.63, 2.75, 2.86, 2.98, 3.10, 3.22,
3.34, 3.45, 3.57, 3.69 (from bottom to top, in $10^{11}$\,cm$^{-2}$). Magnetic field $B_\parallel
= 0$\,T.}
\label{Fig.2}
\end{figure}

Figure \ref{Fig.1} clearly shows  linear regions  in the $\rho(T)$
 dependences at low temperatures. Low resistivity
values indicate that the density range under investigation
belongs to the metallic region ($\rho_D \ll h/e^2$) and the
conditions of the applicability of the interaction corrections
theory \cite{ZNA1} are satisfied. Figure~\ref{Fig.2} shows a
comparison of the conductivity
(recalculated from the  low-temperature data shown in Fig.~\ref{Fig.1})
with interaction corrections \cite{ZNA1}:
\begin{equation}
\sigma(T) = \sigma_D + \delta\sigma_C + 15 \delta\sigma_T,
\label{sigTB0}
\end{equation}
\noindent
where $\sigma(T)$
is the Drude conductivity;
$\delta\sigma_C$ and $\delta\sigma_T$ are the
singlet and triplet terms of the interaction correction,
respectively; and the factor 15 appears due to the two-fold
degeneracy in spin and valleys \cite{delta_v}. The correction
associated with weak localization is small in this temperature
range \cite{pudalov1} and is disregarded in the calculation.

The Fermi-liquid constant $F_0^\sigma$ entering into the triplet
term $\delta\sigma_T$ was measured independently \cite{pudalov2}. The
Drude conductivity $\sigma_D$ was determined by extrapolating
the experimental $\sigma(T)$ data
 to $T= 0$\,K according
to theoretical dependence (\ref{sigTB0}) for the ballistic regime;
$\tau $ was determined
from $\sigma_D$
using the band mass $m^* = 0.205m_e$
(where $m_e$ is the electron mass) \cite{pudalov2}.

It is seen in Fig.~\ref{Fig.2} that, as well as in previous works
\cite{aleiner,savchenko,shashkin}, the calculated interaction corrections quantitatively
describe the low-temperature regions of the
experimental $\sigma(T)$ curves with no free parameters over
the whole studied range of electron densities.

\subsection{Measurements of $\sigma(T)$ in a strong magnetic field}
\begin{figure}[h]
\centerline{\includegraphics[scale=0.9]{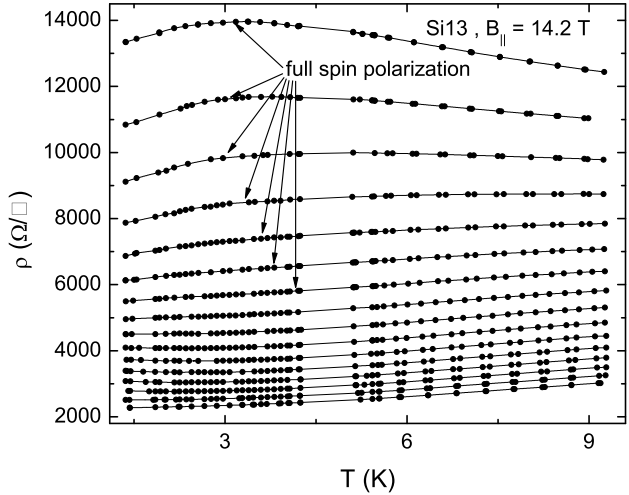}}
\caption{Temperature dependences of $\rho(T)$ for carrier densities
1.92, 2.04, 2.16, 2.28, 2.39, 2.51, 2.63, 2.75, 2.86,
2.98, 3.10, 3.22, 3.34, 3.45, 3.57, 3.69 (from top to bottom, in $10^{11}$\,cm$^{-2}$).
Magnetic field $B_\parallel = 14.2$\,T. The arrows
mark the curves for the full spin polarization of the 2D system
(for densities
1.92, 2.04, 2.16, 2.28, 2.39, 2.51, 2.63).}
\label{Fig.3}
\end{figure}

In order to compare the transport properties of the
system in the spin-polarized state and in zero magnetic
field, we measured the $\rho(T)$ dependences in the in-plane magnetic
field $B_\parallel = 14.2$\,T for the same electron densities as
in zero magnetic field. The experimental dependences
of $\rho(T)$ for various electron densities in magnetic field
$B_\parallel = 14.2$\,T are shown in Fig.~\ref{Fig.3}.

First, it is seen that the temperature dependences
$\rho(T,B_\parallel=14.2$\,T) are substantially weaker than those for zero
magnetic field (Fig.~\ref{Fig.1}). It is also seen that the magnetic
field increases the effective disorder in the system,
which is manifested in the factor of four increase in resistivity
with respect to the data shown in Fig.~\ref{Fig.1}. This fourfold
increase in resistivity agrees with the theory of
temperature-dependent screening \cite{DG00,DSH_SP}, which
attributes it to an increase in $k_F$ by a factor of $\sqrt{2}$, as
well as to a decrease in the density of electron states and to
the factor of two increase in the screening radius.

\begin{figure}[h]
\centerline{\includegraphics[scale=0.95]{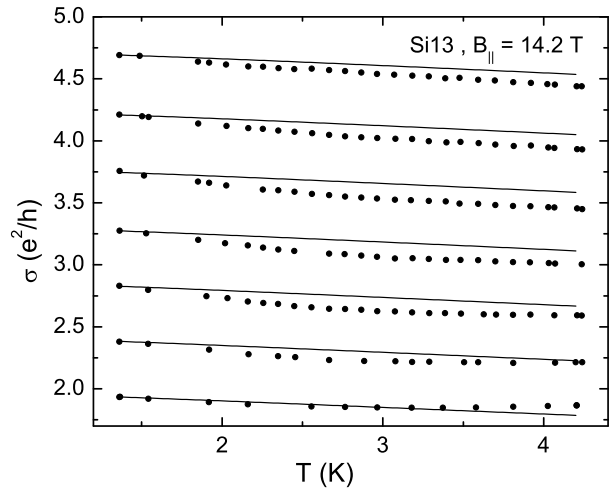}}
\caption{Comparison of the experimental $\sigma(T)$ dependences (dots)
and  calculated interaction corrections (solid
lines) (\protect\ref{sigTB14}) for the densities (from bottom to top)
1.92, 2.04, 2.16, 2.28, 2.39, 2.51, 2.63 (in units of $10^{11}$\,cm$^{-2}$).
Magnetic field $B_\parallel = 14.2$\,T.}
\label{Fig.4}
\end{figure}

Second, the increase in the temperature range of
measurements enabled to reveal that the $\rho(T)$ dependences
in a strong parallel magnetic field are nonmonotonic
(see Fig.~\ref{Fig.3}). Despite the
$\rho(T)$ dependences, in general, are nonlinear  (see Fig.~\ref{Fig.3}),
one can clearly see that for all densities at low-temperatures
they have  linear regions, which can
be compared to the interaction corrections \cite{ZNA1}. The $\rho(T)$
curves  marked with arrows in Fig.~\ref{Fig.3} correspond
to the condition of full spin polarization $g\mu_BB_\parallel > 2E_F$.
The low-temperature regions  of these curves recalculated
to conductivity $\sigma(T)$ are shown in Fig.~\ref{Fig.4} in comparison
with the interaction corrections calculated for the spin-polarized
state \cite{aleiner_private}:

\begin{equation}
\sigma(T) = \sigma_D + \delta\sigma_C + 3\delta\sigma_T,
\label{sigTB14}
\end{equation}

A decrease in the number of the triplet terms from
$(2n_v)^2-1=15$ to $n_v^2-1=3$ is caused by the removal
of spin degeneracy at the unchanged twofold valley degeneracy
$n_v = 2$ \cite{delta_v}. In the calculations of interaction corrections
for the spin-polarized state, we used the same $F_0^\sigma$
and $m^*$ values as those for the case of zero magnetic field \cite{chi_B,shashkin_m}.
In view of an increase in disorder in the system for
the spin-polarized state, we redetermined the $\tau$ and $\sigma_D$
values by extrapolating the $\sigma(T,B_\parallel=14.2$\,T) curves  to
$T = 0$\,K. We emphasize that, although $\tau$ values decreased,
the temperature range of measurements remained to correspond
to the ballistic interaction regime.

It is seen from Fig.~\ref{Fig.4} that the experimental data qualitatively
agree with interaction corrections. Although the
temperature dependences $\sigma(T)$ in the spin-polarized
state are noticeably (a factor of three to five) weaker than those for zero
magnetic field (see Fig.~\ref{Fig.2}), their slopes $d\sigma/dT$ are nonzero,
which contradicts the predictions of the non Fermi-liquid two-phase model
\cite{spivak}. For the lowest densities
(lower lines in Fig.~\ref{Fig.4}), the discrepancy between the
measured and theoretical slopes approximately amounts to
$20–-30$\%. Such a discrepancy appears probably because
the interaction corrections were calculated in Ref.~\cite{ZNA1} for the
region $\sigma_D h/e^2 \gg 1$, whereas $\sigma_D h/e^2 \sim 2-3$ for the lowest
curves in Fig.~\ref{Fig.4}. For higher densities (upper lines in Fig.~\ref{Fig.4}),
the slopes of the experimental lines differ from the calculated
values by a factor of $2-2.8$. The increase in the discrepancy
with electron density is probably associated
with a decrease in the spin polarization degree at finite
temperature: $|g\mu_B B_\parallel - 2E_F| \sim T$. On the other hand,
the theory of temperature-dependent screening \cite{GD86,DG00} predicts
that $(d\sigma(T,B_{pol})/dT)/(d\sigma(T,0)/dT)=1/2(\sigma_D(B_{pol})/\sigma_D(0))=1/8$;
however, according to the experimental data (see
Figs.~\ref{Fig.2},\ref{Fig.4}), this ratio varies only slightly for various densities
and is equal to $\approx 0.04$, i.e., approximately
one third of the predicted value.

Thus, in this work we measured the temperature dependence of the conductivity
of the 2D electron system in Si, both in the absence of a magnetic field and in the
spin-polarized state. It has been found that the temperature
dependence of conductivity in the spin-polarized
state is nonmonotonic, which is not explained by any
available theories. For low temperatures (but in the ballistic
interaction regime, $T\tau > 1$), the temperature
dependence of conductivity is approximately linear
both in zero magnetic field and in the field of full spin polarization;
this enables one to compare the
experimental data with theoretical predictions. The
temperature dependence of the conductivity in the spin-polarized
state is much weaker than that for the case of
zero magnetic field, but it does not vanish. This result
contradicts the two-phase model \cite{spivak}. The $\sigma(T)$ dependences
measured in zero magnetic field are quantitatively
described by the theory of interaction corrections \cite{ZNA1}, whereas in the spin polarized state
the agreement with theory is only qualitative. For both cases, the comparison
with the theory is performed with no free parameters.
The measured ratio $(d\sigma(T,B_{pol})/dT)/(d\sigma(T,0)/dT)$ of the
slopes of the temperature dependences in the spin-polarizing
and zero magnetic fields is larger than the values
predicted by the interaction corrections theory by a factor of
$\approx 1.2-3$  and is about one third of the value predicted by the
theory of temperature-dependent screening.

The work was supported by the grants  from RFBR, INTAS,
Russian Academy of Sciences,  the Presidential Program for the
support of Young Russian researchers
and Leading scientific schools. D.A.K.
acknowledges the support of the Educational-–Scientific
Center, Lebedev Physical Institute,
and Russian Science Support Foundation.

\end{multicols}

\begin{references}
\bibitem{delta_v}
The intervalley splitting is
$\Delta_v \ll T$ over the entire temperature
range, and therefore does not
affect the number of triplet terms in Eqs.~(\protect\ref{sigTB0})
and (\protect\ref{sigTB14}).
\bibitem{chi_B}
In Refs. \protect\cite{aleiner,shashkin_m} it was shown  that the effective mass Si-MOS
structures is independent
of the parallel magnetic field.
The $F_0^\sigma(B_\parallel)$ dependence on $B_\parallel$ is
manifested only in low magnetic fields; in the field of
full spin polarization $F_0^\sigma$  is extrapolated to the same
value as in zero magnetic field \protect\cite{review2}.
\end{references}
\end{document}